\documentclass[english]{revtex4}
\usepackage[T1]{fontenc}
\usepackage[latin1]{inputenc}
\usepackage{float}
\usepackage{graphicx}

\begin{document}
\begin{flushright}JLAB-05-04-305 \\
 December 20, 2004\end{flushright}

\begin{flushright}\vspace{2cm}\end{flushright}

\title{Probing Generalized Parton Distributions with Neutrino Beams%
\footnote{Talk given at Fermilab Proton Driver Workshop, October 6-9 2004.%
}}

\author{A. PSAKER}

\affiliation{Physics Department, Old Dominion University,\\
 Norfolk, VA 23529, USA\\
 and\\
 Theory Group, Jefferson Laboratory,\\
 Newport News, VA 23606, USA\\
 }

\begin{abstract}
\vspace{2cm}

A short review of form factors, parton distribution functions and
generalized parton distributions is given. A possible application
of generalized parton distributions in the weak sector is discussed.

\vspace{5mm}

PACS number(s): 13.15.+g, 13.40.-f, 13.40.Gp, 13.60.-r, 13.60.Fz,
13.60.Hb 
\end{abstract}
\maketitle
\newpage

\section{Introduction}

Specific aspects of hadronic structure are described by several different
phenomenological functions. Form factors, usual parton distribution
functions (PDFs) and distribution amplitudes are the so-called {}``old''
phenomenological functions since they have been around for a long
time. On the other hand, the concept of generalized parton distributions
(GPDs) \cite{Muller:1998fv,Ji:1996ek,Ji:1996nm,Radyushkin:1996nd,Radyushkin:1997ki}
(for recent reviews, see \cite{Goeke:2001tz,Diehl:2003ny}) is new.
These {}``new'' phenomenological functions are hybrids of the {}``old''
ones, and therefore they provide a unified and more detailed description
of the hadronic structure.\\

In recent years, significant effort was made to access GPDs through
the measurement of hard exclusive electro-production processes. The
simplest process in this respect is the deeply virtual Compton scattering
(DVCS) process, i.e. an electron scatters off a nucleon producing
a photon in the final state. Using the neutrino beam instead we extend
the DVCS process into the weak sector, where one expects to be sensitive
to a different flavor decomposition of GPDs, and further, due to the
presence of the axial part of the \emph{V-A} interaction also sensitive
to a different set of GPDs. In addition, the weak DVCS process allows
us to study flavor non-diagonal GPDs, e.g. in the neutron-to-proton
transition. \\

Detailed study of weak deeply virtual Compton scattering with electron
and neutrino beams will be presented in a forthcoming paper.

\section{Phenomenological Functions}

Among phenomenological functions we discuss form factors, usual parton
distribution functions and generalized parton distributions.

\subsection{Form Factors}

Form factors are defined through matrix elements of electromagnetic
and weak currents between the hadronic states. In particular, the
matrix element of the vector current between the nucleon states $N\left(p_{1},s_{1}\right)$
and $N\left(p_{2},s_{2}\right)$ is parametrized in terms of the nucleon
electromagnetic form factors, i.e. Dirac and Pauli form factors,

\begin{eqnarray}
\left\langle N\left(p_{2},s_{2}\right)\right|J_{V}^{\mu}\left(0\right)\left|N\left(p_{1},s_{1}\right)\right\rangle  & = & \overline{u}\left(p_{2},s_{2}\right)\left[\gamma^{\mu}F_{1}\left(t\right)-i\sigma^{\mu\nu}\frac{r_{\nu}}{2M}F_{2}\left(t\right)\right]u\left(p_{1},s_{1}\right),\label{eq:formfactors1}\end{eqnarray}
 where $r=p_{1}-p_{2}$ is the overall momentum transfer, the invariant
$t=r^{2}$ and \emph{M} denotes the nucleon mass. In the case of the
axial vector current one has the axial and pseudoscalar form factors,

\begin{eqnarray}
\left\langle N\left(p_{2},s_{2}\right)\right|J_{A}^{\mu}\left(0\right)\left|N\left(p_{1},s_{1}\right)\right\rangle  & = & \overline{u}\left(p_{2},s_{2}\right)\left[\gamma^{\mu}\gamma_{5}g_{A}\left(t\right)-\gamma_{5}\frac{r^{\mu}}{2M}g_{P}\left(t\right)\right]u\left(p_{1},s_{1}\right).\label{eq:formfactors2}\end{eqnarray}
 Both currents in Eqs. (\ref{eq:formfactors1}) and (\ref{eq:formfactors2})
are given by the sum of their flavor components, $J_{V}^{\mu}\left(0\right)=\sum_{f}Q_{f}\bar{\psi}_{f}\left(0\right)\gamma^{\mu}\psi_{f}\left(0\right)$
and $J_{A}^{\mu}\left(0\right)=\sum_{f}Q_{Af}\bar{\psi}_{f}\left(0\right)\gamma^{\mu}\gamma_{5}\psi_{f}\left(0\right)$,
where $Q_{f}$ and $Q_{Af}$ are the electric (in units of $\left|e\right|$)
and axial charges of the quark of flavor \emph{f}, respectively. A
similar decomposition holds for the form factors, $F_{1,2}\left(t\right)=\sum_{f}Q_{f}F_{1,2f}\left(t\right)$
and $g_{A,P}\left(t\right)=\sum_{f}Q_{Af}g_{A,Pf}\left(t\right)$.
Their limiting values at $t=0$ are known, e.g. Dirac and Pauli form
factors give the total electric charge of the nucleon and its anomalous
magnetic moment. \\

\begin{figure}[H]
\begin{center}

\includegraphics[%
  scale=0.6]{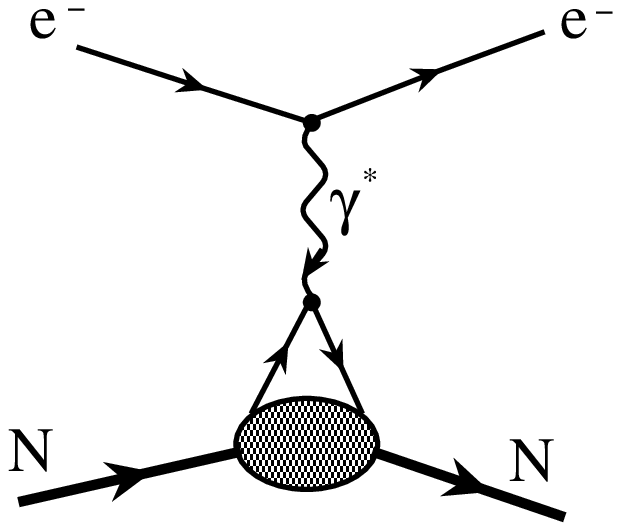} \includegraphics[%
  scale=0.6]{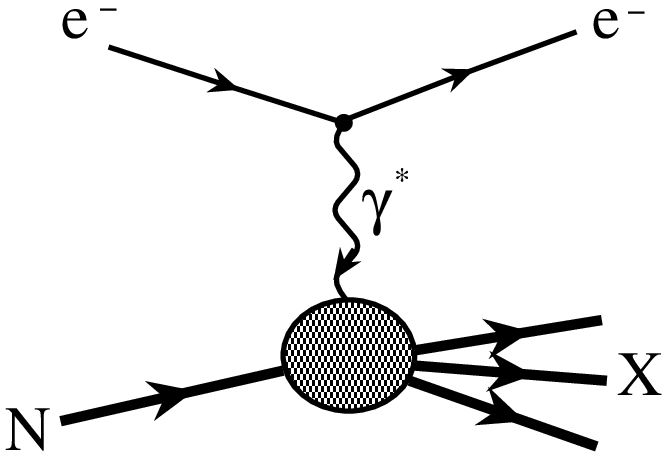}

\end{center}\caption{Elastic electron-nucleon scattering in one-photon exchange approximation (left) and deeply inelastic scattering (right).}

\label{ffanddis}
\end{figure}

In particular, the nucleon electromagnetic form factors can be measured
through elastic electron-nucleon scattering,

\begin{eqnarray}
e^{-}N & \longrightarrow & e^{-}N.\label{eq:elasticscatteringreaction}\end{eqnarray}
 The process is shown in one photon-exchange approximation in Figure
\ref{ffanddis}, left.

\subsection{Usual Parton Distribution Functions }

PDFs are defined through forward matrix elements of quark/gluon fields
separated by light-like distances. For the unpolarized case one has

\begin{eqnarray}
\left\langle N\left(p_{1},s_{1}\right)\right|\bar{\psi}_{f}\left(-z/2\right)\gamma^{\mu}\psi_{f}\left(z/2\right)\left|N\left(p_{1},s_{1}\right)\right\rangle _{z^{2}=0} & = & \overline{u}\left(p_{1},s_{1}\right)\gamma^{\mu}u\left(p_{1},s_{1}\right)\nonumber \\
 &  & \times\int_{0}^{1}dx\;\left[e^{-ix\left(p_{1}z\right)}q_{f}\left(x\right)-e^{ix\left(p_{1}z\right)}\overline{q}_{f}\left(x\right)\right],\label{eq:unpolPDFs}\end{eqnarray}
 and for the polarized one

\begin{eqnarray}
\left\langle N\left(p_{1},s_{1}\right)\right|\bar{\psi}_{f}\left(-z/2\right)\gamma^{\mu}\gamma_{5}\psi_{f}\left(z/2\right)\left|N\left(p_{1},s_{1}\right)\right\rangle _{z^{2}=0} & = & \overline{u}\left(p_{1},s_{1}\right)\gamma^{\mu}\gamma_{5}u\left(p_{1},s_{1}\right)\nonumber \\
 &  & \times\int_{0}^{1}dx\;\left[e^{-ix\left(p_{1}z\right)}\Delta q_{f}\left(x\right)+e^{ix\left(p_{1}z\right)}\Delta\overline{q}_{f}\left(x\right)\right].\label{eq:polPDFs}\end{eqnarray}
 To make connection with GPDs, which are usually discussed in the
region $-1\leq x\leq1$, it is convenient to introduce new distribution
functions,

\begin{equation}
\widetilde{q}_{f}\left(x\right)=\left\{ \begin{array}{cc}
q_{f}\left(x\right) & x>0\\
-\overline{q}_{f}\left(-x\right) & x<0\\
\end{array}\right.\label{eq:newfunctionunpol}\end{equation}
 and

\begin{equation}
\Delta\widetilde{q}_{f}\left(x\right)=\left\{ \begin{array}{cc}
\Delta q_{f}\left(x\right) & x>0\\
\Delta\overline{q}_{f}\left(-x\right) & x<0\\
\end{array}\right.\label{eq:newfunctionpol}\end{equation}
 and alternatively, write the integrals over \emph{x} in Eqs. (\ref{eq:unpolPDFs})
and (\ref{eq:polPDFs}) as

\begin{eqnarray}
\int_{0}^{1}dx\;\left[e^{-ix\left(p_{1}z\right)}q_{f}\left(x\right)-e^{ix\left(p_{1}z\right)}\overline{q}_{f}\left(x\right)\right] & = & \int_{-1}^{1}dx\; e^{-ix\left(p_{1}z\right)}\widetilde{q}_{f}\left(x\right),\nonumber \\
\int_{0}^{1}dx\;\left[e^{-ix\left(p_{1}z\right)}\Delta q_{f}\left(x\right)+e^{ix\left(p_{1}z\right)}\Delta\overline{q}_{f}\left(x\right)\right] & = & \int_{-1}^{1}dx\; e^{-ix\left(p_{1}z\right)}\Delta\widetilde{q}_{f}\left(x\right).\label{eq:newintegrals}\end{eqnarray}
 Furthermore, one observes that the definition of PDFs has the form
of the plane wave decomposition. Thus it allows us to give the momentum
space interpretation: $q_{f}\left(x\right)/\overline{q}_{f}\left(x\right)$
is the probability to find the quark/antiquark of flavor \emph{f}
carrying the momentum $xp_{1}$ inside a fast-moving nucleon having
the momentum $p_{1}$.\\

\begin{figure}[H]
\begin{center}

\includegraphics[%
  scale=0.6]{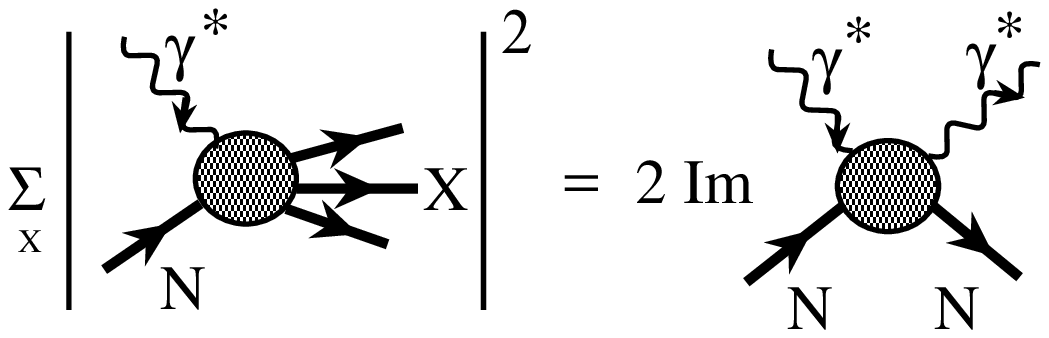}

\end{center}\caption{Optical theorem.}

\label{optical}
\end{figure}

PDFs have been intensively studied in hard inclusive processes for
the last three decades. The classic example in this respect is the
deeply inelastic scattering (DIS) process, i.e. inclusive scattering
of high energy leptons on the nucleon,

\begin{eqnarray}
e^{-}N & \longrightarrow & e^{-}X,\label{eq:disreaction}\end{eqnarray}
 shown in Figure \ref{ffanddis}, right. It played a key role in revealing
the quark structure of the nucleon. The structure functions, accessed
in the DIS process, are directly expressed in terms of PDFs. Through
the optical theorem, its cross section is given by the imaginary part
of the forward virtual Compton scattering amplitude (see Figure \ref{optical}).
The summation over \emph{X} reflects the inclusive nature of the nucleon
structure description by PDFs. \\

\begin{figure}[H]
\begin{center}

\includegraphics[%
  scale=0.6]{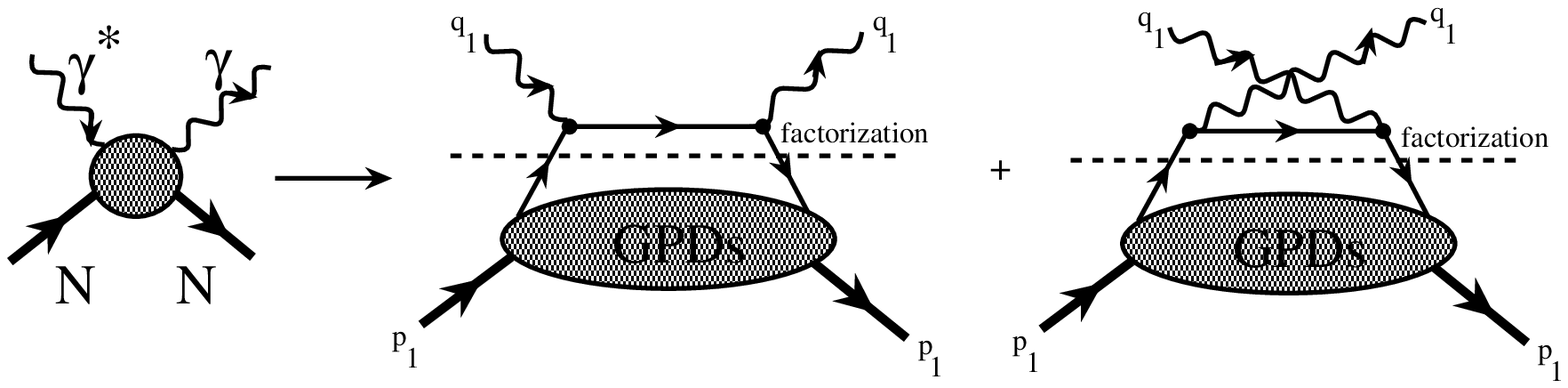}

\end{center}\caption{Forward virtual Compton scattering amplitude.}

\label{fvca}
\end{figure}

In the Bjorken regime, where the space-like momentum transfer is sufficiently
large together with large total center-of-mass energy of the photon-nucleon
system, $-q_{1}^{2}$ and $\left(p_{1}+q_{1}\right)^{2}\rightarrow\infty$,
while the ratio $x_{B}\equiv-q_{1}^{2}/2\left(p_{1}\cdot q_{1}\right)$
is finite, perturbative QCD factorization works. In other words, the
forward virtual Compton scattering amplitude factorizes into perturbatively
calculable hard scattering process at the level of quarks and gluons,
and process independent matrix elements which contain the soft non-perturbative
information about the nucleon structure represented by the blob (see
Figure \ref{fvca}). We recall that these forward matrix elements
consist of quark and gluon operators, whose fields are separated by
a light-like distance. They are described and parametrized in terms
of PDFs. Schematically, QCD factorization allows us to write the amplitude
in the form of the so-called handbag diagrams. Moreover, the leading
contribution in the lowest order in the strong coupling constant $\alpha_{s}$
is given by two (\emph{s}- and \emph{u}-channel) handbag diagrams
in which the hard propagator is convoluted with the PDFs. Taking the
imaginary part of the forward virtual Compton scattering amplitude
generates the delta function, $\Im\left[\left(xp_{1}\pm q_{1}\right)^{2}+i\epsilon\right]^{-1}\approx\delta\left(x\mp x_{B}\right)/2\left(p_{1}\cdot q_{1}\right)$.
Hence in the DIS process one measures the PDF $\widetilde{q}_{f}\left(x\right)$
at two points, $x=\pm x_{B}$, with $x=x_{B}$ corresponding to the
quark PDF and $x=-x_{B}$ for that of antiquarks.

\subsection{Generalized Parton Distributions}

A more recent attempt to use perturbative QCD to extract new information
about the hadronic structure is the study of hard exclusive electro-production
processes, in particular the DVCS process. This is a much more difficult
task due to the small cross sections. However, high energy and high
luminosity electron accelerators combined with large acceptance spectrometers
give a unique opportunity to perform precision studies of such reactions.
The DVCS process can be accessed through the reaction

\begin{eqnarray}
e^{-}N & \longrightarrow & e^{-}N\gamma.\label{eq:DVCSreaction}\end{eqnarray}

\begin{figure}[H]
\begin{center}

\includegraphics[%
  scale=0.6]{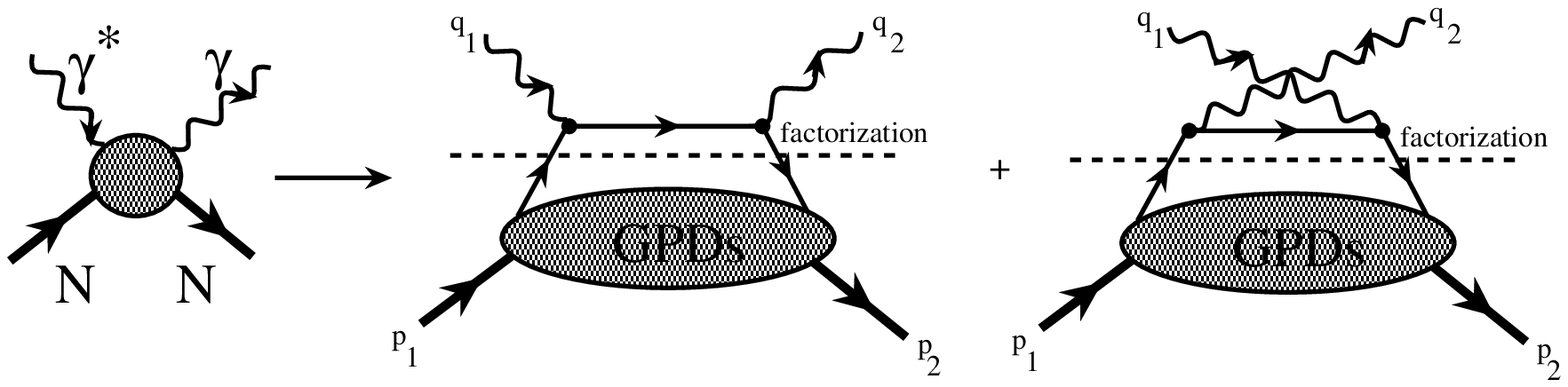}

\end{center}\caption{Non-forward virtual Compton scattering amplitude.}

\label{nonfvca}
\end{figure}

It turns out that factorization into short and long distance dynamics
is more general. Having large space-like virtuality of the initial
photon while the final state photon is on shell is sufficient for
QCD factorization to work. In contrast to the DIS process, the outgoing
photon is real, and henceforth, the overall momentum transfer is not
equal to zero. In the leading handbag approximation, the so-called
non-forward virtual Compton scattering amplitude is dominated by two
diagrams (see Figure \ref{nonfvca}). The lower blob now contains
the non-forward matrix elements of the same quark and gluon operators
as in the forward case. They are parametrized in terms of GPDs.\\

\begin{figure}[H]
\begin{center}

\includegraphics[%
  scale=0.6]{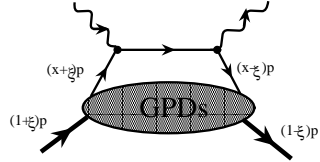}

\end{center}\caption{Symmetric description of the s-channel handbag diagram.}

\label{symmetric}
\end{figure}

It is convenient to introduce the average of the nucleon momenta,
$p=\left(p_{1}+p_{2}\right)/2$, and treat the initial and final hadron
in a symmetric way (see Figure \ref{symmetric}). In this scheme,
at the leading twist-2 level, the nucleon structure information can
be parametrized in terms of two unpolarized and two polarized GPDs
denoted by $H,\; E\;\mathrm{and}\;\widetilde{H},\;\widetilde{E}$,
respectively. They are functions of three variables $\left(x,\xi,t\right)$,
and further they are defined for each quark flavor \emph{f}. In addition
to the usual light-cone momentum fraction \emph{x}, GPDs also depend
upon another scaling variable, the skewness parameter $\xi=r_{\Vert}/2p_{\Vert}$,
specifying the longitudinal momentum asymmetry, and upon the invariant
\emph{t}. The variables \emph{x} and $\xi$ solely characterize the
longitudinal momenta of the partons involved, however, the \emph{t}-dependence
of GPDs is related to their transverse momenta. Thus one can simultaneously
access the longitudinal momentum and transverse position of the parton
in the infinite momentum frame \cite{Burkardt:2000za}. Furthermore,
by removing the quark with the light-cone momentum fraction $x+\xi$
and replacing it with the quark of the momentum fraction $x-\xi$,
one can say that GPDs measure the coherence between two different
quark momentum states of the nucleon, i.e. the quark momentum correlations
in the nucleon, whereas usual PDFs yield only the probability that
a quark carries a fraction \emph{x} of the nucleon momentum.\\

Since $-1\leq x\leq1$ in Figure \ref{symmetric}, the momentum fractions
$x\pm\xi$ of the active quarks can be either positive or negative.
Positive and negative momentum fractions corresponds to quarks and
antiquarks, respectively. Therefore GPDs have three distinct regions:
when $\xi\leq x\leq1\;\left(-1\leq x\leq-\xi\right)$ both partons
represent quarks (antiquarks), whereas for $-\xi\leq x\leq\xi$ one
parton represents a quark, and the other parton an antiquark. In the
first two regions GPDs are just the generalizations of the usual PDFs,
however, in the third region they behave like a meson distribution
amplitude. Hence, in the region $-\xi\leq x\leq\xi$, they contain
new information about the nucleon structure since this region is not
present in DIS. \\

GPDs have interesting properties linking them to usual PDFs and form
factors. In the forward limit, $p_{1}=p_{2}$ and $\xi,\; r,\; t=0$,
the GPDs $H\;\mathrm{and}\;\widetilde{H}$ coincide with the quark
density distribution $q_{f}\left(x\right)$ and the quark helicity
distribution $\Delta q_{f}\left(x\right)$ given by Eqs. (\ref{eq:newfunctionunpol})
and (\ref{eq:newfunctionpol}) obtained from the DIS process. One
writes the so-called reduction formulas for the functions $H\;\mathrm{and}\;\widetilde{H}$,

\begin{equation}
H_{f}\left(x,0,0\right)=\left\{ \begin{array}{cc}
q_{f}\left(x\right) & x>0\\
-\overline{q}_{f}\left(-x\right) & x<0\\
\end{array}\right.\label{eq:forwardlimit1}\end{equation}
 and

\begin{equation}
\widetilde{H}_{f}\left(x,0,0\right)=\left\{ \begin{array}{cc}
\Delta q_{f}\left(x\right) & x>0\\
\Delta\overline{q}_{f}\left(-x\right) & x<0\\
\end{array}\right.\label{eq:forwardlimit2}\end{equation}
 while the functions $E\;\mathrm{and}\;\widetilde{E}$ have no connections
to PDFs. They are always accompanied with the momentum transfer \emph{r},
and therefore invisible in inclusive measurements. In the local limit,
$z=0$, GPDs reduce to the form factors. In other words, the first
moments of GPDs are equal to the nucleon elastic form factors. Namely,

\begin{eqnarray}
\int_{-1}^{1}dx\; H_{f}\left(x,\xi,t\right)=F_{1f}\left(t\right) & , & \int_{-1}^{1}dx\; E_{f}\left(x,\xi,t\right)=F_{2f}\left(t\right),\nonumber \\
\int_{-1}^{1}dx\;\widetilde{H}_{f}\left(x,\xi,t\right)=g_{Af}\left(t\right) & , & \int_{-1}^{1}dx\;\widetilde{E}_{f}\left(x,\xi,t\right)=g_{Pf}\left(t\right).\label{eq:sumrules}\end{eqnarray}
 We call these relations the sum rules. They are model and $\xi$-independent.\\

GPDs are also relevant for the nucleon spin structure. In particular,
the second moment of the unpolarized GPDs at $t=0$ gives the quark
angular momentum,

\begin{eqnarray}
J_{q} & = & \frac{1}{2}\sum_{f}\int_{-1}^{1}dx\; x\left[H_{f}\left(x,\xi,t=0\right)+E_{f}\left(x,\xi,t=0\right)\right].\label{eq:quarkangularmomentum}\end{eqnarray}
 The above equation is independent of $\xi$. The quark angular momentum,
on the other hand, decomposes into the quark intrinsic spin and the
quark orbital angular momentum,

\begin{eqnarray}
J_{q} & = & \frac{1}{2}\Delta\Sigma+L_{q},\label{eq:quarkangularmomentumdecomposition}\end{eqnarray}
 where $\Delta\Sigma$ is measured through the polarized DIS process.
Substituting Eq. (\ref{eq:quarkangularmomentum}) into Eq. (\ref{eq:quarkangularmomentumdecomposition})
one can determine $L_{q}$. Moreover, since the nucleon spin comes
from quarks and gluons, $1/2=J_{q}+J_{g}$, one can extract the gluon
contribution $J_{g}$ to the nucleon spin. Hence by measuring GPDs
one obtains information about the angular momentum distributions of
quarks and gluons in the hadron.

\section{Weak DVCS Amplitude}

\begin{figure}[H]
\begin{center}

\includegraphics[%
  scale=0.6]{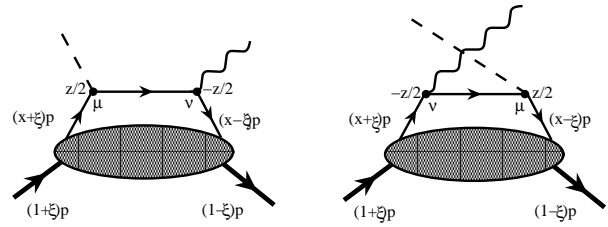}

\end{center}\caption{Handbag diagrams (s- and u-channel) in the weak DVCS amplitude.}

\label{handbagdiagrams}
\end{figure}

In the most general case, the virtual Compton scattering amplitude
is given by a Fourier transform of the correlation function of two
electroweak currents. In particular, for the weak DVCS process we
have

\begin{eqnarray}
T^{\mu\nu} & = & i\int d^{4}z\; e^{-iqz}\left\langle N\left(p-r/2,s_{2}\right)\right|T\left\{ J_{W}^{\mu}\left(z/2\right)J_{EM}^{\nu}\left(-z/2\right)\right\} \left|N\left(p+r/2,s_{1}\right)\right\rangle ,\label{eq:weakamplitude}\end{eqnarray}
 where $J_{W}^{\mu}\left(z/2\right)$ corresponds to either the weak
neutral current $J_{WN}^{\mu}\left(z/2\right)$ or to the weak charged
current $J_{WC}^{\mu}\left(z/2\right)$, i.e. to the exchange of the
weak boson $Z^{0}$ or $W^{\pm}$, respectively. One of the methods
to study the behavior of Eq. (\ref{eq:weakamplitude}) in the generalized
Bjorken region is to use the light-cone expansion for the time-ordered
product of two currents $T\left\{ J_{W}^{\mu}\left(z/2\right)J_{EM}^{\nu}\left(-z/2\right)\right\} $
in the coordinate representation. The expansion is performed in terms
of QCD string operators. Its leading order contribution is shown in
Figure \ref{handbagdiagrams}. The hard part of both handbag diagrams
starts at the zeroth order in $\alpha_{s}$ with the purely tree level
diagrams in which the weak virtual boson and real photon interact
with quarks being treated as massless fermions. Since the weak current
couples to the quark current through two types of vertices, $qqZ^{0}$
and $qqW^{\pm}$, the quark fields at coordinates $\pm z/2$ can carry
either the same or different flavor quantum numbers. We treat these
two cases separately.

\subsection{Weak Neutral Current}

We expand the time-ordered product of the weak neutral and electromagnetic
current in Eq. (\ref{eq:weakamplitude}),

\begin{eqnarray}
iT\left\{ J_{WN}^{\mu}\left(z/2\right)J_{EM}^{\nu}\left(-z/2\right)\right\}  & = & -\frac{z_{\rho}}{4\pi^{2}z^{4}}\sum_{f}Q_{f}\left[\bar{\psi}_{f}\left(-z/2\right)\gamma^{\nu}\gamma^{\rho}\gamma^{\mu}\left(c_{V}^{f}-\gamma_{5}c_{A}^{f}\right)\psi_{f}\left(z/2\right)\right.\nonumber \\
 &  & \left.-\bar{\psi}_{f}\left(z/2\right)\gamma^{\mu}\left(c_{V}^{f}-\gamma_{5}c_{A}^{f}\right)\gamma^{\rho}\gamma^{\nu}\psi_{f}\left(-z/2\right)\right],\label{eq:weakneutralexpansion1}\end{eqnarray}
 where $c_{V}^{f}$ and $c_{A}^{f}$ are the weak vector and axial
vector charges, respectively. We express the original bilocal quark
operators with three Lorentz indices in Eq. (\ref{eq:weakneutralexpansion1})
in terms of the string operators with only one Lorentz index,

\begin{eqnarray}
iT\left\{ J_{WN}^{\mu}\left(z/2\right)J_{EM}^{\nu}\left(-z/2\right)\right\}  & = & -\frac{z_{\rho}}{4\pi^{2}z^{4}}\sum_{f}Q_{f}\left\{ c_{V}^{f}s^{\mu\rho\nu\eta}\left[\bar{\psi}_{f}\left(-z/2\right)\gamma_{\eta}\psi_{f}\left(z/2\right)-\left(z\longrightarrow-z\right)\right]\right.\nonumber \\
 &  & +c_{V}^{f}i\epsilon^{\mu\rho\nu\eta}\left[\bar{\psi}_{f}\left(-z/2\right)\gamma_{\eta}\gamma_{5}\psi_{f}\left(z/2\right)+\left(z\longrightarrow-z\right)\right]\nonumber \\
 &  & -c_{A}^{f}s^{\mu\rho\nu\eta}\left[\bar{\psi}_{f}\left(-z/2\right)\gamma_{\eta}\gamma_{5}\psi_{f}\left(z/2\right)-\left(z\longrightarrow-z\right)\right]\nonumber \\
 &  & \left.-c_{A}^{f}i\epsilon^{\mu\rho\nu\eta}\left[\bar{\psi}_{f}\left(-z/2\right)\gamma_{\eta}\psi_{f}\left(z/2\right)+\left(z\longrightarrow-z\right)\right]\right\} .\label{eq:weakneutralexpansion2}\end{eqnarray}
 We have two types of string operators in Eq. (\ref{eq:weakneutralexpansion2}).
The (axial) vector string operators come (with) without $\gamma_{5}$.
They can be accompanied with tensors, $s^{\mu\rho\nu\eta}$ and $\epsilon^{\mu\rho\nu\eta}$,
which are symmetric or antisymmetric in indices $\mu,\;\nu$. Furthermore,
in addition to the standard electromagnetic DVCS process, we end up
with two more terms.\\

In the next step we isolate the twist-2 part of the string operators
in Eq. (\ref{eq:weakneutralexpansion2}), and sandwich it between
the initial and final nucleon state. At this point we introduce the
relevant non-perturbative functions (GPDs) in order to parametrize
the non-forward nucleon matrix elements of the vector and axial vector
string operators on the light-cone. Namely,

\begin{eqnarray}
\left\langle N\left(p_{2},s_{2}\right)\right|\bar{\psi}_{f}\left(-z/2\right)\not\! z\psi_{f}\left(z/2\right)\pm\left(z\longrightarrow-z\right)\left|N\left(p_{1},s_{1}\right)\right\rangle _{z^{2}=0} & = & \overline{u}(p_{2},s_{2})\not\! zu(p_{1},s_{1})\int_{-1}^{1}dx\; e^{-ix\left(pz\right)}H_{f}^{\pm}\left(x,\xi,t\right)\nonumber \\
 &  & +\overline{u}(p_{2},s_{2})\frac{\left(\not\! z\not\! r-\not\! r\not\! z\right)}{4M}u(p_{1},s_{1})\nonumber \\
 &  & \times\int_{-1}^{1}dx\; e^{-ix\left(pz\right)}E_{f}^{\pm}\left(x,\xi,t\right),\nonumber \\
\left\langle N\left(p_{2},s_{2}\right)\right|\bar{\psi}_{f}\left(-z/2\right)\not\! z\gamma_{5}\psi_{f}\left(z/2\right)\mp\left(z\longrightarrow-z\right)\left|N\left(p_{1},s_{1}\right)\right\rangle _{z^{2}=0} & = & \overline{u}(p_{2},s_{2})\not\! z\gamma_{5}u(p_{1},s_{1})\int_{-1}^{1}dx\; e^{-ix\left(pz\right)}\widetilde{H}_{f}^{\pm}\left(x,\xi,t\right)\nonumber \\
 &  & -\overline{u}(p_{2},s_{2})\frac{\left(r\cdot z\right)}{2M}\gamma_{5}u(p_{1},s_{1})\nonumber \\
 &  & \times\int_{-1}^{1}dx\; e^{-ix\left(pz\right)}\widetilde{E}_{f}^{\pm}\left(x,\xi,t\right).\label{eq:parametrization}\end{eqnarray}
 The {}``plus'' distributions enter the electromagnetic DVCS process,
and they correspond to the sum of quark and antiquark contributions,
i.e. to the sum of contributions from valence quarks and twice the
sea quarks. On the other hand, the weak version of the process gives
access to the {}``minus'' GPDs, which correspond to the difference
of quark and antiquark contributions. This difference is equal to
the valence quark contribution.

\subsection{Weak Charged Current}

The expansion of the time-ordered product of two currents in the weak
charged sector reads

\begin{eqnarray}
iT\left\{ J_{WC}^{\mu}\left(z/2\right)J_{EM}^{\nu}\left(-z/2\right)\right\}  & = & -\frac{z_{\rho}}{4\pi^{2}z^{4}}\sum_{f,f'}\left[Q_{f'}\bar{\psi}_{f'}\left(-z/2\right)\gamma^{\nu}\gamma^{\rho}\gamma^{\mu}\left(1-\gamma_{5}\right)\psi_{f}\left(z/2\right)\right.\nonumber \\
 &  & \left.-Q_{f}\bar{\psi}_{f}\left(z/2\right)\gamma^{\mu}\left(1-\gamma_{5}\right)\gamma^{\rho}\gamma^{\nu}\psi_{f'}\left(-z/2\right)\right].\label{eq:weakchargedexpansion}\end{eqnarray}
 Here the sum over quark flavors is subject to an extra condition,
$Q_{f}-Q_{f'}=1$ or $-1$, due the fact that the weak virtual boson
$W^{\pm}$ carries an electric charge. Hence the initial and final
nucleon are not the same particles anymore. Now the vector and axial
vector string operators obtained from Eq. (\ref{eq:weakchargedexpansion})
are accompanied by different electric charges and quark flavors. For
that reason the non-forward nucleon matrix elements,

\begin{eqnarray}
\left\langle N\left(p_{2},s_{2}\right)\right|\widehat{O}^{f'f}\left(z\right)\left|N\left(p_{1},s_{1}\right)\right\rangle  & = & \left\langle N\left(p_{2},s_{2}\right)\right|\bar{\psi}_{f'}\left(-z/2\right)\not\! z\psi_{f}\left(z/2\right)\pm\left(f'\longrightarrow f,z\longrightarrow-z\right)\left|N\left(p_{1},s_{1}\right)\right\rangle ,\nonumber \\
\left\langle N\left(p_{2},s_{2}\right)\right|\widehat{O}_{5}^{f'f}\left(z\right)\left|N\left(p_{1},s_{1}\right)\right\rangle  & = & \left\langle N\left(p_{2},s_{2}\right)\right|\bar{\psi}_{f'}\left(-z/2\right)\not\! z\gamma_{5}\psi_{f}\left(z/2\right)\mp\left(f'\longrightarrow f,z\longrightarrow-z\right)\left|N\left(p_{1},s_{1}\right)\right\rangle ,\label{eq:flavornondiagonalmatrixelements}\end{eqnarray}
 involve different flavor combinations. They are parametrized in terms
of GPDs, which are non-diagonal in quark flavor,

\begin{eqnarray}
\left\langle N\left(p_{2},s_{2}\right)\right|\widehat{O}^{f'f}\left(z\right)\left|N\left(p_{1},s_{1}\right)\right\rangle _{z^{2}=0} & = & \overline{u}(p_{2},s_{2})\not\! zu(p_{1},s_{1})\int_{-1}^{1}dx\; e^{-ix\left(pz\right)}H_{f'f}^{\pm}\left(x,\xi,t\right)\nonumber \\
 &  & +\overline{u}(p_{2},s_{2})\frac{\left(\not\! z\not\! r-\not\! r\not\! z\right)}{4M}u(p_{1},s_{1})\int_{-1}^{1}dx\; e^{-ix\left(pz\right)}E_{f'f}^{\pm}\left(x,\xi,t\right),\nonumber \\
\left\langle N\left(p_{2},s_{2}\right)\right|\widehat{O}_{5}^{f'f}\left(z\right)\left|N\left(p_{1},s_{1}\right)\right\rangle _{z^{2}=0} & = & \overline{u}(p_{2},s_{2})\not\! z\gamma_{5}u(p_{1},s_{1})\int_{-1}^{1}dx\; e^{-ix\left(pz\right)}\widetilde{H}_{f'f}^{\pm}\left(x,\xi,t\right)\nonumber \\
 &  & -\overline{u}(p_{2},s_{2})\frac{\left(r\cdot z\right)}{2M}\gamma_{5}u(p_{1},s_{1})\int_{-1}^{1}dx\; e^{-ix\left(pz\right)}\widetilde{E}_{f'f}^{\pm}\left(x,\xi,t\right).\label{eq:nondiagonalparametrization}\end{eqnarray}

\section{Weak DVCS Processes}

Before we examine specific processes we introduce a simple model and
discuss the kinematics which is common to all DVCS-like reactions.\\

Our simple model has three properties. First we assume that the sea
quark contribution is negligible, and therefore the {}``plus'' GPDs
are equal to the {}``minus'' GPDs with the quark flavor $f=u,d$.
Secondly we take a factorized ansatz of the \emph{t}-dependence from
the other two scaling variables \emph{x} and $\xi$ for all distributions.
The \emph{t}-dependence of GPDs is characterized by the corresponding
form factors given by Eq. (\ref{eq:sumrules}). Thirdly we neglect
the $\xi$-dependence in all GPDs except in the $\widetilde{E}$ distribution.
The parametrization of GPDs is taken from Refs. \cite{Guichon:1998xv,Radyushkin:1998rt,Belitsky:2001ns,Goshtasbpour:1995eh,Penttinen:1999th}.
Namely, for the \emph{H} and \emph{E} distributions one has \cite{Guichon:1998xv}

\begin{eqnarray}
H_{u}^{val}\left(x,\xi,t\right)=q_{u}^{val}\left(x\right)\frac{F_{1u}\left(t\right)}{2} & \mathrm{and} & H_{d}^{val}\left(x,\xi,t\right)=q_{d}^{val}\left(x\right)F_{1d}\left(t\right),\nonumber \\
E_{u}^{val}\left(x,\xi,t\right)=q_{u}^{val}\left(x\right)\frac{F_{2u}\left(t\right)}{2} & \mathrm{and} & E_{d}^{val}\left(x,\xi,t\right)=q_{d}^{val}\left(x\right)F_{2d}\left(t\right),\label{eq:HandE}\end{eqnarray}
 where the unpolarized valence quark distributions in the proton are
given by \cite{Radyushkin:1998rt}

\begin{eqnarray}
q_{u}^{val}\left(x\right)=1.89x^{-0.4}\left(1-x\right)^{3.5}\left(1+6x\right) & \mathrm{and} & q_{d}^{val}\left(x\right)=0.54x^{-0.6}\left(1-x\right)^{4.2}\left(1+8x\right).\label{eq:unpolvalencedistributions}\end{eqnarray}
 The \emph{u}- and \emph{d}-quark form factors can be extracted from
the proton and neutron form factors, which can further be related
to the Sachs electric and magnetic form factors. For the polarized
quark GPD $\widetilde{H}$ one has \cite{Belitsky:2001ns}

\begin{eqnarray}
\widetilde{H}_{u}^{val}\left(x,\xi,t\right)=\Delta q_{u}^{val}\left(x\right)\left(1+\frac{\left|t\right|}{m_{A}^{2}}\right)^{-2} & \mathrm{and} & \widetilde{H}_{d}^{val}\left(x,\xi,t\right)=\Delta q_{d}^{val}\left(x\right)\left(1+\frac{\left|t\right|}{m_{A}^{2}}\right)^{-2},\label{eq:Htilda}\end{eqnarray}
 with the mass $m_{A}=1.03\;\mathrm{GeV}$. The polarized valence
quark distributions can be expressed in terms of the unpolarized ones
through \cite{Goshtasbpour:1995eh}

\begin{eqnarray}
\Delta q_{u}^{val}=\cos\theta_{D}\left(q_{u}^{val}-\frac{2}{3}q_{d}^{val}\right) & \mathrm{and} & \Delta q_{d}^{val}=\cos\theta_{D}\left(-\frac{1}{3}q_{d}^{val}\right),\label{eq:polvalencedistributions}\end{eqnarray}
 where $\cos\theta_{D}=\left[1+\mathrm{H}_{0}\left(1-x^{2}\right)/\sqrt{x}\right]^{-1}$
and $\mathrm{H}_{0}=0.06$. Finally, for $\widetilde{E}$ we accept
the pion pole dominated ansatz \cite{Penttinen:1999th},

\begin{eqnarray}
\widetilde{E}_{u}^{val}\left(x,\xi,t\right)=\frac{1}{2}F_{\pi}\left(t\right)\frac{\theta\left(\left|x\right|<\xi\right)}{2\xi}\;\phi_{\pi}\left(\frac{x+\xi}{2\xi}\right) & \mathrm{and} & \widetilde{E}_{d}^{val}\left(x,\xi,t\right)=-\widetilde{E}_{u}^{val}\left(x,\xi,t\right).\label{eq:Etilda}\end{eqnarray}
 The function $F_{\pi}\left(t\right)$ and the pion distribution amplitude
$\phi_{\pi}$ are taken in the form

\begin{eqnarray}
F_{\pi}\left(t\right)=4g_{A}\left(t=0\right)M^{2}\left[\frac{1}{\left(m_{\pi}^{2}+\left|t\right|\right)/\mathrm{GeV^{2}}}-\frac{1.7}{\left(1+\left|t\right|/2\;\mathrm{GeV^{2}}\right)^{2}}\right] & \mathrm{and} & \phi_{\pi}\left(u\right)=6u\left(1-u\right),\label{eq:Fpiandphipi}\end{eqnarray}
 where $m_{\pi}$ denotes the pion mass and $g_{A}\left(t=0\right)=1.267$.\\

\begin{figure}[H]
\begin{center}

\includegraphics[%
  scale=0.6]{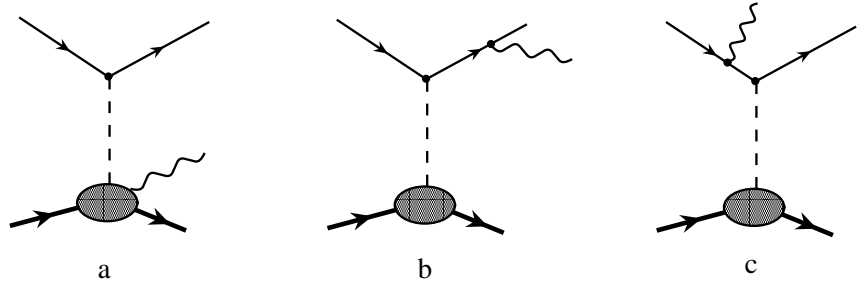}

\end{center}\caption{DVCS (a) and Bethe-Heitler (b and c) diagrams contributing to the lepton-production of a real photon.}

\label{weakprocess}
\end{figure}

In general, a neutrino induced DVCS process on a nucleon is given
by the reaction

\begin{eqnarray}
l_{1}\left(k\right)+N_{1}\left(p_{1}\right) & \longrightarrow & l_{2}\left(k'\right)+N_{2}\left(p_{2}\right)+\gamma\left(q_{2}\right),\label{eq:weakDVCSreaction}\end{eqnarray}
 where a neutrino $l_{1}$ scatters from a nucleon $N_{1}$ to a final
state $l_{2}$, nucleon $N_{2}$ and real photon $\gamma$. Schematically,
the reaction is presented by three diagrams (see Figure \ref{weakprocess}).
The first diagram is the so-called DVCS diagram which corresponds
to the emission of the real photon from the nucleon blob. In our approximation
it is calculated from two handbag diagrams (see Figure \ref{handbagdiagrams}).
In the other two (Bethe-Heitler) diagrams of Figure \ref{weakprocess}
the real photon is emitted from a lepton leg.\\

\begin{figure}[H]
\begin{center}

\includegraphics[%
  scale=0.6]{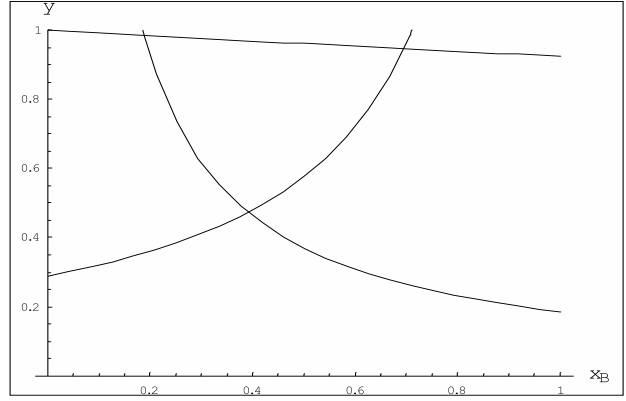}

\end{center}\caption{Kinematically allowed region.}

\label{xyplane}
\end{figure}

\begin{figure}[H]
\begin{center}

\includegraphics[%
  scale=0.6]{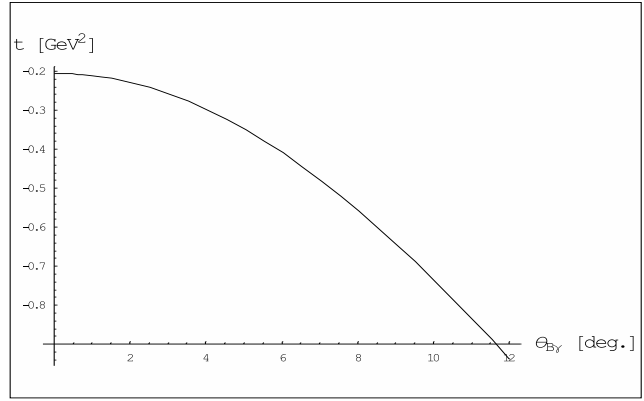}

\end{center}\caption{Momentum transfer squared plotted as a function of the angle between the incoming weak virtual boson and outgoing real photon in the target rest frame.}

\label{t}
\end{figure}

The differential cross section in the target rest frame, in which
the weak virtual boson four-momentum $q_{1}$ has no transverse components,
assumes the form

\begin{eqnarray}
\frac{d^{4}\sigma}{dx_{B}dQ_{1}^{2}dtd\varphi} & = & \frac{1}{64s}\frac{1}{\left(2\pi\right)^{4}}\frac{1+x_{B}\left(M/\omega\right)}{M\omega x_{B}\left(y+2x_{B}\left(M/\omega\right)\right)^{2}}\left|T\right|^{2},\label{eq:diffcrosssection}\end{eqnarray}
 where \emph{T} represents the invariant matrix element. The invariants
in Eq. (\ref{eq:diffcrosssection}) are given by $s\equiv\left(k+p_{1}\right)^{2},\; Q_{1}^{2}\equiv-q_{1}^{2},\;\mathrm{and}\; y\equiv\left(p_{1}\cdot q_{1}\right)/\left(p_{1}\cdot k\right)$.
Moreover, $\omega$ denotes the energy of the incoming neutrino beam,
and $\varphi$ the angle between the lepton and nucleon scattering
planes. One finds the kinematically allowed region for the reaction
(see Figure \ref{xyplane}) under the following constraints: fixed
neutrino beam energy at $\omega=5.75\;\mathrm{GeV}$, the invariant
mass squared of the weak virtual boson-nucleon system, $\hat{s}\equiv\left(p_{1}+q_{1}\right)^{2}\geq4\;\mathrm{GeV^{2}}$,
and the virtuality of the weak boson, $Q_{1}^{2}\geq2\;\mathrm{GeV^{2}}$.
As an example of the particular kinematics, we further choose $Q_{1}^{2}=2.5\;\mathrm{GeV}^{2},\; x_{B}=0.4$
and set $\varphi=0$. One plots \emph{t} as a function of the angle
$\theta_{B\gamma}$ between the incoming weak virtual boson \emph{B}
and the outgoing real photon $\gamma$ in the target rest frame (see
Figure \ref{t}).\\

\begin{figure}[H]
\begin{center}

\includegraphics[%
  scale=0.6]{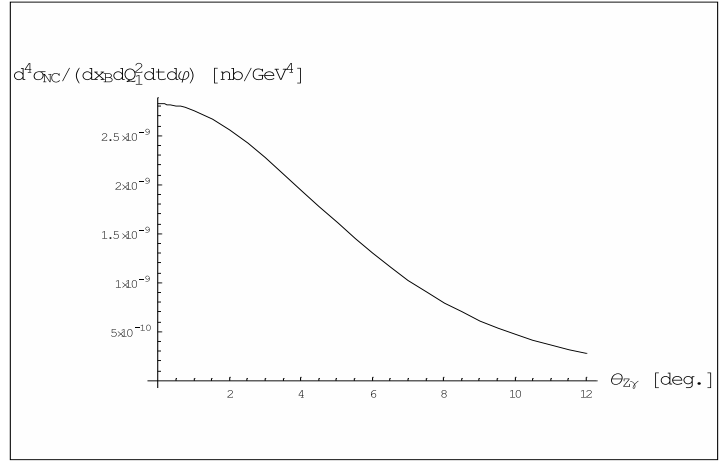}

\end{center}\caption{Weak neutral DVCS differential cross section on an unpolarized proton target plotted as a function of the angle between the incoming weak virtual boson and outgoing real photon in the target rest frame.}

\label{weakneutral}
\end{figure}

\begin{figure}[H]
\begin{center}

\includegraphics[%
  scale=0.6]{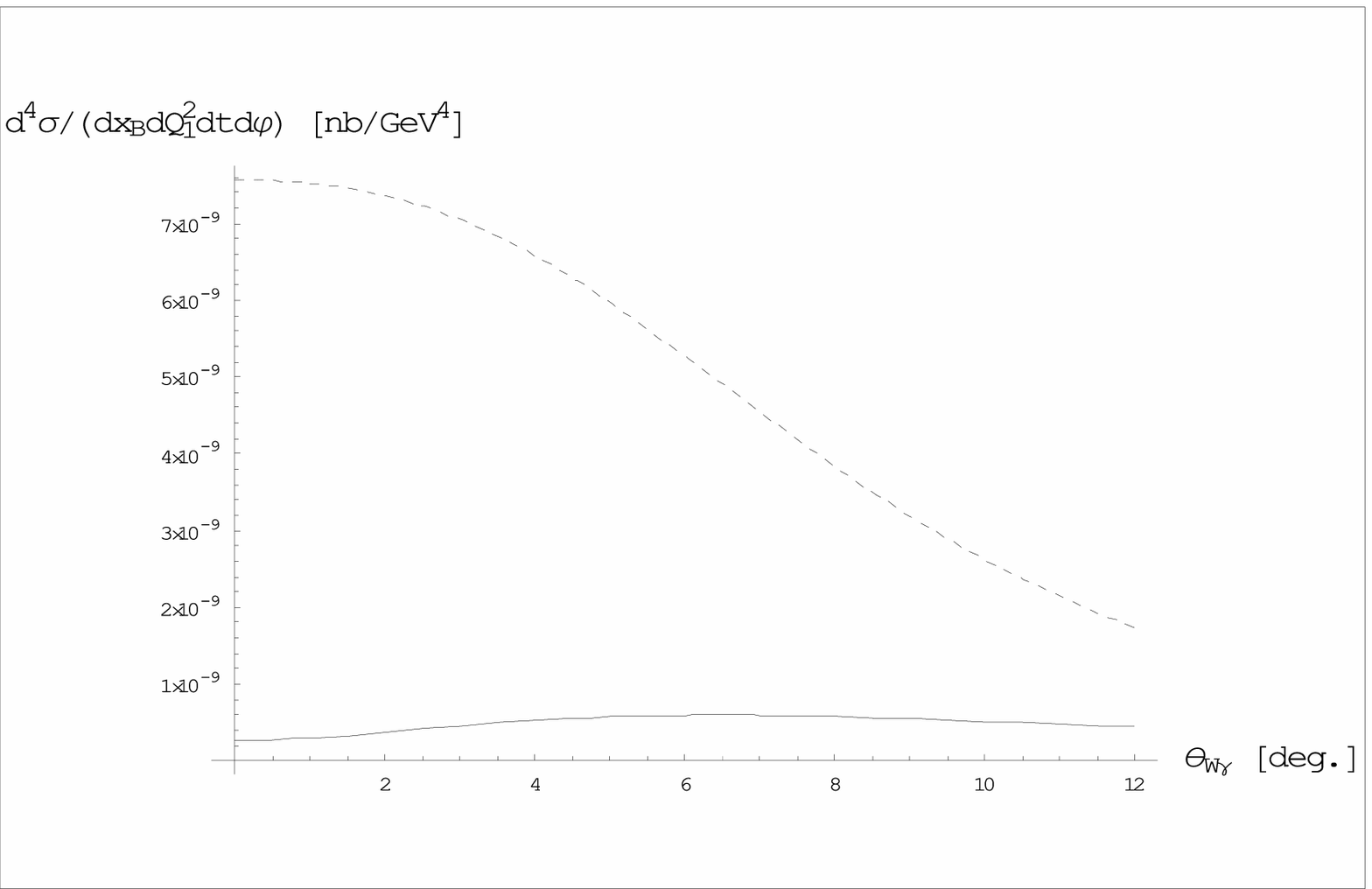}

\end{center}\caption{Compton (dashed line) and Bethe-Heitler (solid line) contribution to the weak charged DVCS differential cross section on an unpolarized neutron target plotted as a function of the angle between the incoming weak virtual boson and outgoing real photon in the target rest frame.}

\label{weakcharged}
\end{figure}

\begin{figure}[H]
\begin{center}

\includegraphics[%
  scale=0.6]{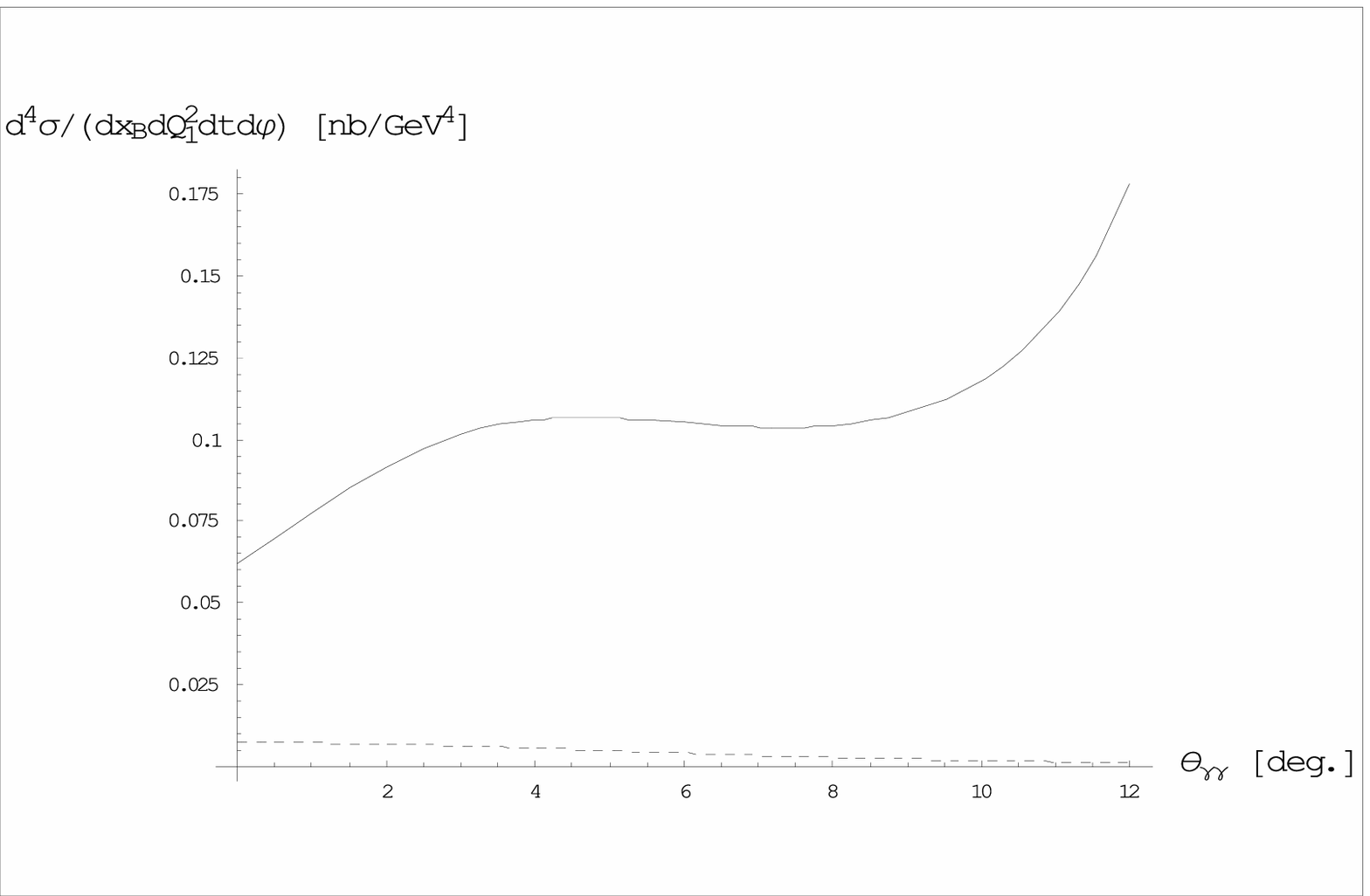}

\end{center}\caption{Compton (dashed line) and Bethe-Heitler (solid line) contribution to the electromagnetic DVCS differential cross section on an unpolarized proton target plotted as a function of the angle between the incoming weak virtual boson and outgoing real photon in the target rest frame.}

\label{standard}
\end{figure}

We study two different examples of the weak DVCS processes. In the
weak neutral sector we consider neutrino scattering off an unpolarized
proton target through the exchange of $Z^{0}$. Here one only measures
the Compton contribution (see Figure \ref{weakneutral}) since the
photon can not be emitted by the neutrino. Next we consider neutrino-neutron
scattering through the exchange of $W^{+}$ with a proton in the final
state. This particular process, however, has both contributions (see
Figure \ref{weakcharged}). The Bethe-Heitler background is calculated
from one diagram (see diagram (b) in Figure \ref{weakprocess}) since
only the outgoing muon can emit the real photon. In contrast to the
standard electromagnetic DVCS process on an unpolarized proton target
(see Figure \ref{standard}), where the Bethe-Heitler cross section
is well above the Compton one, the situation in the weak charged sector
is the other way round.

\section{Conclusions}

Generalized parton distributions contain the most complete and unified
description of the internal quark-gluon structure of hadrons. Form
factors, usual parton distribution functions and distribution amplitudes,
on the other hand, can be treated just as particular or limiting cases
of generalized parton distributions. Furthermore, the formalism of
generalized parton distributions provides nontrivial relations between
exclusive and inclusive processes, and also between different exclusive
processes.\\

We have extended the deeply virtual Compton scattering process into
the weak sector by using the neutrino beam. We have argued that the
weak deeply virtual Compton scattering process gives an additional
information about the hadronic structure. It is expected in the near
future that neutrino scattering off a nucleon will be studied at high
intensity neutrino beam facilities.

\begin{acknowledgments}
I would like to thank my advisor Prof. A. Radyushkin and W. Melnitchouk
for their useful comments. This work was supported by the US Department
of Energy DE-FG02-97ER41028 and by the contract DE-AC05-84ER40150
under which the Southeastern Universities Research Association (SURA)
operates the Thomas Jefferson Accelerator Facility.
\end{acknowledgments}

\section{Figures and Plots}

\begin{thebibliography}{10}
\bibitem{Muller:1998fv}D.~Muller, D.~Robaschik, B.~Geyer, F.~M.~Dittes and J.~Horejsi,
Fortsch.\ Phys.\ {} \textbf{42}, 101 (1994) {[}arXiv:hep-ph/9812448{]}. 
\bibitem{Ji:1996ek}X.~D.~Ji, Phys.\ Rev.\ Lett.\ {} \textbf{78}, 610 (1997) {[}arXiv:hep-ph/9603249{]}. 
\bibitem{Ji:1996nm}X.~D.~Ji, Phys.\ Rev.\ D \textbf{55}, 7114 (1997) {[}arXiv:hep-ph/9609381{]}. 
\bibitem{Radyushkin:1996nd}A.~V.~Radyushkin, Phys.\ Lett.\ B \textbf{380}, 417 (1996) {[}arXiv:hep-ph/9604317{]}. 
\bibitem{Radyushkin:1997ki}A.~V.~Radyushkin, Phys.\ Rev.\ D \textbf{56}, 5524 (1997) {[}arXiv:hep-ph/9704207{]}. 
\bibitem{Goeke:2001tz}K.~Goeke, M.~V.~Polyakov and M.~Vanderhaeghen, Prog.\ Part.\ Nucl.\ Phys.\ {}
\textbf{47}, 401 (2001) {[}arXiv:hep-ph/0106012{]}. 
\bibitem{Diehl:2003ny}M.~Diehl, Phys.\ Rept.\ {} \textbf{388}, 41 (2003) {[}arXiv:hep-ph/0307382{]}. 
\bibitem{Burkardt:2000za}M.~Burkardt, Phys.\ Rev.\ D \textbf{62}, 071503 (2000) {[}Erratum-ibid.\ D
\textbf{66}, 119903 (2002){]}{[}arXiv:hep-ph/0005108{]}. 
\bibitem{Guichon:1998xv}P.~A.~M.~Guichon and M.~Vanderhaeghen, Prog.\ Part.\ Nucl.\ Phys.\ {}
\textbf{41}, 125 (1998) {[}arXiv:hep-ph/9806305{]}. 
\bibitem{Radyushkin:1998rt}A.~V.~Radyushkin, Phys.\ Rev.\ D \textbf{58}, 114008 (1998) {[}arXiv:hep-ph/9803316{]}. 
\bibitem{Belitsky:2001ns}A.~V.~Belitsky, D.~Muller and A.~Kirchner, Nucl.\ Phys.\ B \textbf{629},
323 (2002) {[}arXiv:hep-ph/0112108{]}. 
\bibitem{Goshtasbpour:1995eh}M.~Goshtasbpour and G.~P.~Ramsey, Phys.\ Rev.\ D \textbf{55},
1244 (1997) {[}arXiv:hep-ph/9512250{]}. 
\bibitem{Penttinen:1999th}M.~Penttinen, M.~V.~Polyakov and K.~Goeke, Phys.\ Rev.\ D \textbf{62},
014024 (2000) {[}arXiv:hep-ph/9909489{]}.
\end{thebibliography}
\end{document}